\documentclass[twocolumn,letter]{jpsj2}
\def\runtitle{DMPK Equation for Transmission Eigenvaluse
in Metallic Carbon Nanotubes}
\def\runauthor{Yositake {\sc Takane}}
\title{%
DMPK Equation for Transmission Eigenvalues in Metallic Carbon Nanotubes
}

\author{%
Yositake {\sc Takane}
}

\inst{%
Department of Quantum Matter, Graduate School of Advanced Sciences of Matter,
Hiroshima University, Higashi-Hiroshima 739-8530, Japan
}

\abst{%
The Dorokhov-Mello-Pereyra-Kumar (DMPK) equation for transmission eigenvalues
is derived for metallic carbon nanotubes with several conducting channels
when the potential range of scatterers is larger than the lattice constant.
With increasing system length $L$,
the system approaches a fixed point, where only one channel is perfectly
conducting and other channels are completely closed.
The asymptotic behavior of the conductance in the long-$L$ regime
is investigated on the basis of the DMPK equation.
It is shown that the length scale for
the exponential decay of the typical conductance
is reduced due to the presence of the perfectly conducting channel.
If a magnetic field is applied, the system falls into the unitary class.
It is pointed out that this transition is characterized by
the disappearance of the perfectly conducting channel and
the increase in decay length for the typical conductance.
}

\begin{document}
\sloppy
\maketitle

Carbon nanotubes (CNs) are micrometer-long hollow cylinders with
nanometer-scale radii.~\cite{iijima}
We can regard CNs as a new class of quasi-one-dimensional quantum wires.
Recent experiments have enabled the electrical transport measurement of
individual single-wall CNs.~\cite{tans,bezryadin,mceuen}
We expect that the unique topological structures of CNs induce
unusual electron transport properties,
which have not been observed in ordinary quantum wire systems.
Ando and Nakanishi~\cite{ando1,nakanishi} studied electron scattering
in CNs in the case where only one conducting channel is present
and showed the absence of backward scattering
when the potential range of scatterers is larger than the lattice constant.

Recently, Ando and Suzuura~\cite{ando2} studied metallic CNs
with long-range impurity potential in the case
where the Fermi level lies in several conducting channels
and showed within a ${\mib k} \cdot {\mib p}$ scheme
that one perfectly conducting channel is present.
They also performed numerical simulations based on a tight-binding model.
Their numerical result suggests that, with increasing system length $L$,
the system approaches a fixed point, where only one channel is perfectly
conducting and other channels are completely closed.
This denotes that the conductance $G$ decreases towards the quantized value
$G_{0}$ governed by the perfectly conducting channel.
Due to the spin and valley multiplicity, the quantized conductance
is given by $G_{0} = 4e^{2}/h$.
Takane and Wakabayashi~\cite{takane} studied the same problem by adopting
a random-matrix theory
and showed that the fixed point is stabilized by the antilocalization effect.
They also showed that $G$ in the presence of electron decoherence
behaves as $G \propto L_{\phi}/L$ when $l \lesssim L_{\phi} \ll L$,
where $l$ and $L_{\phi}$ are the mean free path and
the phase coherence length, respectively.
These results indicate that the transport properties of CNs
with several conducting channels are very different from
those observed in ordinary quantum wire systems.
However, our understanding of this problem
is not sufficient at the present stage.
Indeed, we do not know how the system approaches the fixed point
with increasing system length.

To answer such a question, we must investigate the scattering problem
in CNs from a statistical viewpoint.
Generally, electron scattering in a quantum wire system with
$N$ channels is described by the scattering matrix $s$,
which is a $2N \times 2N$ matrix given by~\cite{beenakker}
\begin{equation}
       s = \left( \begin{array}{cc}
                    r & t' \\
                    t & r' \\
                  \end{array}
           \right) ,
\end{equation}
with $N \times N$ reflection matrices $r$ and $r'$
(reflection from left to left and from right to right)
and transmission matrices $t$ and $t'$
(transmission from left to right and from right to left).
The Hermitian matrices $t'{t'}^{\dagger}$ and $t t^{\dagger}$
have the same set of eigenvalues $T_{1}, T_{2}, \cdots, T_{N}$.
The transport properties of a sample are determined by $\{T_{a}\}$.
For example, the conductance is given by $G=(2e^{2}/h)\Gamma$ with
$\Gamma = \sum_{a=1}^{N}T_{a}$,
where the factor two corresponds to the electron spin.
In some cases, it is convenient to use $\lambda_{a} \equiv (1-T_{a})/T_{a}$
instead of $T_{a}$.
Let us call $T_{a}$ and $\lambda_{a}$ the transmission eigenvalues.
If the probability distribution of the transmission eigenvalues
is obtained as a function of $L$,
we can completely describe the statistical properties of $G$.
In ordinary quantum wire systems,
the evolution of the distribution function with increasing $L$ is
described by the Dorokhov-Mello-Pereyra-Kumar (DMPK)
equation.~\cite{beenakker,dorokhov,mello1}
We can obtain useful information from it
in considering the statistical properties of $G$.

In this letter, we derive the DMPK equation for the transmission eigenvalues
in metallic CNs with long-range impurity potential
and, on the basis of the resulting DMPK equation,
investigate the asymptotic behavior of the conductance in the regime
where the system length $L$ is much longer than the mean free path $l$.
In deriving the DMPK equation, we adopt the random-matrix model
presented by Takane and Wakabayashi~\cite{takane}
and consider the unique scattering symmetry observed in CNs.
With increasing $L$,
the conductance decreases towards the quantized value $G_{0}$.
We calculate several statistical averages which characterize
the asymptotic behavior of the conductance.
It is shown that the length scale for the exponential decay
of the typical conductance is reduced
due to the presence of the perfectly conducting channel.
If a magnetic field is applied, the system falls into the unitary class.
We point out that this transition is characterized by the disappearance of
the perfectly conducting channel and the increase in decay length.

The number of conducting channels, $N$, in a metallic CN is odd,
so we set $N = 2m+1$ ($m$: integer).
Ando and Suzuura~\cite{ando2} studied the scattering problem
when the potential range of scatterers is larger than the lattice constant
and obtained the symmetry relations for the elements of $s$.
Their result reads
\begin{align}
     \label{eq:symmetry1}
 ^{t}t & = t' ,
       \\
     \label{eq:symmetry2}
 ^{t}r & = - r \quad {\rm and} \quad ^{t}r' = - r' .
\end{align}
Equation~(\ref{eq:symmetry1}) holds for arbitrary systems
with the time-reversal symmetry,
while eq.~(\ref{eq:symmetry2}) characterizes the peculiarity of our system.
Equation~(\ref{eq:symmetry2}) directly results in the absence of
backscattering, $r_{aa} = 0$.~\cite{ando1,nakanishi}
Due to the valley multiplicity, the conductance
is given by $G = 4(e^{2}/h) \Gamma$.
We introduce the transfer matrix $M$, which can be expressed in terms of
the elements of $s$,
\begin{equation}
     \label{eq:m-s}
       M = \left( \begin{array}{cc}
                    (t^{\dagger})^{-1} & r'{t'}^{-1} \\
                    -{t'}^{-1}r & {t'}^{-1} \\
                  \end{array}
           \right) .
\end{equation}
We employ the parameterization~\cite{takane,mello2}
\begin{equation}
     \label{eq:dm}
       M = \left( \begin{array}{cc}
                    {\rm e}^{\theta} & 0 \\
                    0 & {\rm e}^{\theta^{*}} \\
                   \end{array}
           \right)
           \left( \begin{array}{cc}
      \left(1 + \eta \eta^{\dagger}\right)^{\frac{1}{2}} & \eta \\
      \eta^{\dagger} & \left(1 + \eta^{\dagger}\eta\right)^{\frac{1}{2}} \\
                   \end{array}
           \right)
\end{equation}
with $\theta = {\rm i}h$, where $h$ is an arbitrary $N \times N$ Hermitian
matrix and $\eta$ is an arbitrary $N \times N$ complex matrix.
They satisfy $\theta^{\dagger} = - \theta$ and $^{t}\eta = - \eta$.
Equation~(\ref{eq:dm}) ensures both
the flux conservation and the unique scattering symmetry.
In terms of $\theta$ and $\eta$, we can express the elements of $s$ as
\begin{align}
  t & = {\rm e}^{\theta} (1+\eta \eta^{\dagger} )^{-\frac{1}{2}} ,
          \\
 \label{eq:t'-para}
  t'& = (1+\eta^{\dagger} \eta )^{-\frac{1}{2}}{\rm e}^{-\theta^{*}} ,
          \\
  r & = - (1+\eta^{\dagger} \eta )^{-\frac{1}{2}} \eta^{\dagger} ,
          \\
  \label{eq:r'-para}
  r'& = {\rm e}^{\theta} \eta (1+\eta^{\dagger} \eta )^{-\frac{1}{2}}
           {\rm e}^{-\theta^{*}} .
\end{align}

We introduce the transfer matrix $M_{1}$ for a system of length $L$,
which is parameterized by $\theta_{1}$ and $\eta_{1}$ with eq.~(\ref{eq:dm}).
Let ${t'}_{1}$ be the corresponding transmission matrix,
and $T_{a}$ and $\lambda_{a}$ be the eigenvalues of
${t'}_{1}{t'}_{1}^{\dagger}$ and $\eta_{1}^{\dagger}\eta_{1}$, respectively.
We see from eq.~(\ref{eq:t'-para}) that $T_{a}=1/(\lambda_{a}+1)$.
Due to the skew-symmetric nature of $\eta_{1}$, one of $\lambda_{a}$ is
equal to $0$ and all the other eigenvalues are twofold degenerate.
This indicates that one of $T_{a}$ is unity,
resulting in the presence of one perfectly conducting channel,~\cite{ando2}
and all the other eigenvalues are twofold degenerate.
This is a peculiar nature of CNs with long-range impurity potential.
We introduce a unitary matrix $v$, reducing ${t'}_{1}{t'}_{1}^{\dagger}$
to the diagonal form by the transformation
\begin{equation}
  v{t'}_{1}{t'}_{1}^{\dagger}v^{\dagger}
  = {\rm diag} \left(T_{1}, T_{2}, \cdots , T_{N} \right) .
\end{equation}
We find that
\begin{equation}
      \label{eq:eta-hat}
  v{t'}_{1}{t'}_{1}^{\dagger}v^{\dagger}
   = (1+ \hat{\eta}_{1}^{\dagger} \hat{\eta}_{1} )^{-1} ,
\end{equation}
where $\hat{\eta}_{1} = v^{*} \eta_{1} v^{\dagger}$.
Note that $^{t}\hat{\eta}_{1} = -\hat{\eta}_{1}$.
Without loss of generality,
we assume that the $N$th channel is perfectly conducting
(i.e., $\lambda_{N} = 0$)
and express the $N \times N$ skew-symmetric matrix $\hat{\eta}_{1}$ as
\begin{equation}
      \label{eq:eta-1}
   \hat{\eta}_{1}
   = \left( \begin{array}{ccc}
               0 & \xi & 0 \\
             -\xi &  0 & \vdots \\
               0  &  \cdots & 0
            \end{array}
     \right) ,
\end{equation}
where $\xi$ is an $m \times m$ diagonal matrix given by
\begin{equation}
      \label{eq:xi-diag}
  \xi = {\rm diag} \left(\sqrt{\lambda_{1}}, \sqrt{\lambda_{2}},
                           \cdots, \sqrt{\lambda_{m}} \right) .
\end{equation}
We easily see that
\begin{equation}
   \hat{\eta}_{1}^{\dagger}\hat{\eta}_{1}
   = {\rm diag} \left( \lambda_{1}, \lambda_{2}, \cdots, \lambda_{m},
                       \lambda_{1}, \lambda_{2}, \cdots, \lambda_{m}, 0
                \right) .
\end{equation}
The above argument leads to $T_{a}=T_{a+m}=1/(1+\lambda_{a})$
for $1 \le a \le m$ and $T_{N}=1$.
We shall use the notation, $\bar{a} \equiv a+m$ for $1 \le a \le m$
and $\bar{a} \equiv a-m$ for $m+1 \le a \le 2m$.
Using $\bar{a}$, we can rewrite the degeneracy relation $T_{a}=T_{a+m}$
($1 \le a \le m$) as $T_{a} = T_{\bar{a}}$ ($1 \le a \le N-1$).

To study the evolution of $T_{a}$ with increasing $L$,
we attach a small segment of length $\delta x$ to the right-hand side
of the system.
If the transfer matrix for the small segment is $\delta M$,
the transfer matrix $M_{2}$ for the combined system is given by
\begin{equation}
    \label{eq:m=mm}
  M_{2} = \delta M M_{1} .
\end{equation}
Let ${t'}_{2}$ be the corresponding transmission matrix.
The eigenvalues $T'_{a}$ of ${t'}_{2}{t'}_{2}^{\dagger}$ are slightly
different from $T_{a}$, so we express $T'_{a} = T_{a}+\delta T_{a}$.
To obtain the DMPK equation for
the distribution function $P(T_{1}, T_{2}, \cdots, T_{m}, L)$,
we must calculate $\langle \delta T_{a} \rangle_{\rm ss}$
and $\langle \delta T_{a} \delta T_{b} \rangle_{\rm ss}$,
where $\langle \cdots \rangle_{\rm ss}$ denotes
the ensemble average with respect to the small segment.
The DMPK equation is given by~\cite{beenakker}
\begin{align}
      \label{eq:dmpk}
 & \frac{\partial P}{\partial L} =
     \delta x^{-1}\sum_{a=1}^{m} \frac{\partial}{\partial T_{a}}
     \bigg(- \langle \delta T_{a} \rangle_{\rm ss} P
              \nonumber \\
 & \hspace{30mm}
           + \frac{1}{2}\sum_{b=1}^{m}\frac{\partial}{\partial T_{b}}
             \langle \delta T_{a} \delta T_{b} \rangle_{\rm ss} P
     \bigg) .
\end{align}
We apply the parameterization given in eq.~(\ref{eq:dm}) to $\delta M$.
Following Mello and Tomsovic,~\cite{mello2} we assume that
$\langle \theta \rangle_{\rm ss} = \langle \eta \rangle_{\rm ss} = 0$,
$\langle \theta_{ab} \eta_{cd} \rangle_{\rm ss} =
\langle \theta_{ab} \eta_{cd}^{\dagger} \rangle_{\rm ss} =
\langle \eta_{ab} \eta_{cd} \rangle_{\rm ss} = 0$,
$\langle \theta^{2} + \eta \eta^{\dagger} \rangle_{\rm ss} = 0$ and
\begin{align}
  \langle \theta_{ab} \theta_{cd} \rangle_{\rm ss}
    & = \kappa_{ab,cd}^{1} , \\
  \langle \theta_{ab} \theta_{cd}^{*} \rangle_{\rm ss}
    & = \kappa_{ab,cd}^{2} , \\
  \langle \eta_{ab} \eta_{cd}^{\dagger} \rangle_{\rm ss}
    & = \kappa_{ab,cd}^{3} .
\end{align}
We shall take the weak-scattering limit,~\cite{mello2}
where a moment higher than the second plays no role,
so that $\delta M$ can be expressed in the simple form
\begin{equation}
     \label{eq:dmm}
   \delta M = \left( \begin{array}{cc}
                       1 + \theta & \eta \\
                       \eta^{\dagger} & 1 + \theta^{*} \\
                     \end{array}
              \right) .
\end{equation}
We can relate ${t'}_{2}$ and ${t'}_{1}$ by eq.~(\ref{eq:m=mm}).
From eqs.~(\ref{eq:m-s}), (\ref{eq:m=mm}) and (\ref{eq:dmm}),
we obtain
\begin{equation}
     \label{eq:delta-t}
  {t'}_{2}
    = {t'}_{1} \left( 1 - \theta^{*} - \eta^{\dagger}{r'}_{1}
                     + (\theta^{*})^{2} \right) .
\end{equation}

We calculate $\langle \delta T_{a} \rangle_{\rm ss}$ and
$\langle \delta T_{a} \delta T_{b} \rangle_{\rm ss}$ by perturbation theory.
To second order in perturbation theory, we obtain
\begin{equation}
     \label{eq:pertur}
  \delta T_{a} = w_{aa} + \sum_{b(\neq a)} \frac{w_{ab}w_{ba}}{T_{a}-T_{b}} ,
\end{equation}
where $w_{ab}$ is an element of the Hermitian matrix
${t'}_{2}{t'}_{2}^{\dagger} - {t'}_{1}{t'}_{1}^{\dagger}$ in the basis
where ${t'}_{1}{t'}_{1}^{\dagger}$ is diagonal.
The Hermitian matrix is given by
\begin{equation}
  w = v{t'}_{1}
              \left(  (\theta^{*})^{2}
                    + \eta^{\dagger}{r'}_{1}{r'}_{1}^{\dagger}\eta
                    - {r'}_{1}^{\dagger} \eta - \eta^{\dagger}{r'}_{1}
              \right){t'}_{1}^{\dagger}v^{\dagger} .
\end{equation}
We here take the weak-scattering limit;~\cite{mello2}
$\delta x$ approaches $0$ and at the same time, we let scattering
in the small segment become infinitely weak,
so that
\begin{equation}
\lim_{\delta x \to 0 \atop \kappa \to 0} (\delta x)^{-1}
          \kappa_{ab,cd}^{i}
 = \sigma_{ab,cd}^{i}
\end{equation}
are finite quantities,
while $(\delta x)^{-1}$ times a moment higher than the second vanishes.
Accordingly, we calculate
\begin{align}
  f_{a} & = \lim_{\delta x \to 0 \atop \kappa \to 0} (\delta x)^{-1}
  \langle \delta T_{a} \rangle_{\rm ss} ,
       \\
  f_{ab} & = \lim_{\delta x \to 0 \atop \kappa \to 0} (\delta x)^{-1}
  \langle \delta T_{a} \delta T_{b}\rangle_{\rm ss} ,
\end{align}
instead of $\langle \delta T_{a} \rangle_{\rm ss}$
and $\langle \delta T_{a} \delta T_{b} \rangle_{\rm ss}$.
In accordance with $\theta^{\dagger} = - \theta$ and $^{t}\eta = - \eta$,
we adopt the model~\cite{takane} in which $\sigma_{ab,cd}^{i}$ is given by
\begin{align}
      \label{eq:sigma1}
  \sigma_{ab,cd}^{1}
  & = \lim_{\delta x \to 0 \atop \kappa \to 0} (\delta x)^{-1}
      \langle \theta_{ab} \theta_{cd} \rangle_{\rm ss}
    = - \delta_{ad}\delta_{bc} \sigma'_{ab} ,
       \\
      \label{eq:sigma2}
  \sigma_{ab,cd}^{2}
  & = \lim_{\delta x \to 0 \atop \kappa \to 0} (\delta x)^{-1}
      \langle \theta_{ab} \theta_{cd}^{*} \rangle_{\rm ss}
    = \delta_{ac}\delta_{bd} \sigma'_{ab} ,
       \\
     \label{eq:origin}
  \sigma_{ab,cd}^{3}
  & = \lim_{\delta x \to 0 \atop \kappa \to 0} (\delta x)^{-1}
      \langle \eta_{ab} \eta_{cd}^{\dagger} \rangle_{\rm ss}
    = \left( \delta_{ad}\delta_{bc} - \delta_{ac}\delta_{bd} \right)
                \sigma_{ab} .
\end{align}
Here, $\sigma_{ab}$ represents the average reflection
coefficient per unit length from $b$ to $a$,
while $\sigma'_{ab}$ represents the average transmission coefficient.
They satisfy $\sum_{b}\sigma_{ab} = \sum_{b}\sigma'_{ab}$
since $\langle \theta^{2} + \eta \eta^{\dagger}\rangle_{\rm ss} = 0$.
To proceed, we must assume that $\sigma_{ab}$ is expressed
in a simple form.
We adopt the simplest choice~\cite{takane}
\begin{equation}
    \label{eq:ECM}
  \sigma_{ab} = \frac{1-\delta_{ab}}{(N-1)l} ,
\end{equation}
where $l$ is the mean free path.
This ensures the absence of backscattering
(i.e., $\sigma_{aa}=0$).~\cite{ando1,nakanishi}
Equation~(\ref{eq:ECM}) leads to
$\sum_{b}\sigma_{ab} = \sum_{b}\sigma'_{ab} = l^{-1}$.

In calculating $f_{a}$,
we substitute the matrix elements of $w$ into eq.~(\ref{eq:pertur})
and then take the weak-scattering limit after the ensemble average
with respect to the small segment.
The ensemble average is carried out based on
eqs.(\ref{eq:sigma1})-(\ref{eq:ECM}).
Employing eqs.~(\ref{eq:t'-para}), (\ref{eq:r'-para})
and (\ref{eq:eta-hat})-(\ref{eq:xi-diag}), we obtain $f_{N}=0$ and
\begin{align}
      \label{eq:fa}
  f_{a}
   & = - \frac{T_{a}}{l}
       + \frac{T_{a}}{(N-1)l}
                \nonumber \\
   & \hspace{10mm} \times
         \Bigg( 1-T_{a} + \sum_{\scriptstyle b=1
         \atop \scriptstyle (b \neq a, \bar{a})}^{N}
         \frac{T_{a}+T_{b}-2T_{a}T_{b}}{T_{a}-T_{b}} \Bigg)
                \nonumber \\
   & = - \frac{T_{a}}{l}
       + \frac{T_{a}}{(N-1)l}
                \nonumber \\
   & \hspace{10mm} \times
         \Bigg( -T_{a} + 2\sum_{\scriptstyle b=1
         \atop \scriptstyle (b \neq a, \bar{a})}^{m}
         \frac{T_{a}+T_{b}-2T_{a}T_{b}}{T_{a}-T_{b}} \Bigg)
\end{align}
for $1 \le a \le N-1$.
Similarly, we obtain $f_{aN}=0$ for $1 \le a \le N$ and
\begin{equation}
      \label{eq:faa}
  f_{ab} = \left( \delta_{a,b} + \delta_{\bar{a},b} \right)
           \frac{2}{(N-1)l} T_{a}^{2} (1-T_{a})
\end{equation}
for $1 \le a, b \le N-1$.
Substituting eqs.~(\ref{eq:fa}) and (\ref{eq:faa}) into eq.~(\ref{eq:dmpk}),
we obtain the evolution equation
\begin{align}
      \label{eq:dmpk-T}
 & \frac{\partial P(T_{1}, \cdots, T_{m},s)}{\partial s}
   = l \sum_{a=1}^{m} \frac{\partial}{\partial T_{a}}
     \bigg(- f_{a}P(T_{1}, \cdots, T_{m},s)
             \nonumber \\
 & \hspace{25mm}
         + \frac{1}{2}\frac{\partial}{\partial T_{a}}f_{aa}
                           P(T_{1}, \cdots, T_{m},s) \bigg) ,
\end{align}
where $s = L/l$ is the normalized system length.
Although we have treated $T_{a}$ as independent variables,
it is more convenient to use $\lambda_{a} = (1-T_{a})/T_{a}$ in some cases.
Upon a change of variables from $T_{a}$ to $\lambda_{a}$,
we obtain the DMPK equation for $P(\lambda_{1}, \cdots, \lambda_{m},s)$,
\begin{align}
      \label{eq:dmpk-lambda}
 & \frac{\partial P(\lambda_{1}, \cdots, \lambda_{m},s)}{\partial s}
   = \frac{1}{N-1}\sum_{a=1}^{m}
           \nonumber \\
 & \hspace{10mm} \times
   \frac{\partial}{\partial \lambda_{a}}
   \left( \lambda_{a}(1-\lambda_{a}) J
          \frac{\partial}{\partial \lambda_{a}}
            \left( \frac{ P(\lambda_{1}, \cdots, \lambda_{m},s) }{J} \right)
   \right) ,
\end{align}
where
\begin{equation}
      \label{eq:jacob}
  J = \prod_{c=1}^{m} \lambda_{c}^{2}
      \times
      \prod_{b=1}^{m-1}\prod_{a=b+1}^{m}|\lambda_{a}-\lambda_{b}|^{4} .
\end{equation}
Equations~(\ref{eq:dmpk-T}) and (\ref{eq:dmpk-lambda}) are
the central results of the present letter.

From eq.~(\ref{eq:dmpk-T}), we can derive the evolution equation for
an arbitrary function $F(T_{1}, \cdots, T_{m})$.~\cite{mello3}
Multiplying both sides of eq.~(\ref{eq:dmpk-T}) by $F$
and integrating over $\{T_{a}\}$, we obtain
\begin{align}
    \label{eq:evolution-F}
  \frac{\partial \langle F \rangle}{\partial s}
 & = \sum_{a=1}^{m}
     \Biggr\langle 
     \Biggl[-T_{a}
               \nonumber \\
 & \hspace{-2mm}
        + \frac{T_{a}}{N-1}
                  \biggl(-T_{a}
                        + 2 \sum_{\scriptstyle b=1
                                \atop \scriptstyle (b \neq a)}^{m}
                         \frac{T_{a}+T_{b}-2T_{a}T_{b}}{T_{a}-T_{b}}
                  \biggr)
     \Biggr] \frac{\partial F}{\partial T_{a}}
     \Biggr\rangle
       \nonumber \\
& \hspace{3mm}
 + \sum_{a=1}^{m}\frac{1}{N-1}
   \Biggr\langle T_{a}^{2}(1-T_{a}) \frac{\partial^{2} F}{\partial T_{a}^{2}}
   \Biggr\rangle ,
\end{align}
where the ensemble average $\langle \cdots \rangle$ is defined by
\begin{equation}
  \langle \cdots \rangle =
  \int \prod_{a=1}^{m}{\rm d}T_{a} \cdots
              P(T_{1}, \cdots, T_{m},s) .
\end{equation}
As an example, we derive the evolution equations for $\Gamma^{p}$,
where $\Gamma = 1 + 2 \sum_{a=1}^{m}T_{a}$.
Substituting $F = \Gamma^{p}$ into eq.~(\ref{eq:evolution-F}),
we obtain
\begin{align}
      \label{eq:tp}
  (N-1) \frac{\partial \langle \Gamma^{p} \rangle}{\partial s}
& = - p
      \langle \Gamma^{p+1} \rangle + p \langle \Gamma_{2}\Gamma^{p-1} \rangle
            \nonumber \\
& \hspace{7mm}
    + 2p(p-1)
      \langle \left(\Gamma_{2} - \Gamma_{3} \right)\Gamma^{p-2} \rangle ,
\end{align}
where $\Gamma_{q}= 1 + 2 \sum_{a=1}^{m}T_{a}^{q}$.
Equation~(\ref{eq:tp}) is in agreement with
the previously reported result.~\cite{takane}

Now we consider the asymptotic behavior of the conductance $G$
in the long-$L$ regime based on the DMPK equation.~\cite{beenakker,dorokhov}
We introduce a new set of variables $x_{a}$, related to $\lambda_{a}$ by
\begin{equation}
  \lambda_{a}=\sinh^{2}x_{a} ,
\end{equation}
where $x_{a} \ge 0$.
The perfectly conducting channel is characterized by $x_{N} = 0$.
If we make a change of variables from $\lambda_{a}$ to $x_{a}$,
the evolution equation for $P(x_{1}, x_{2}, \cdots, x_{m}, s)$ becomes
\begin{equation}
      \label{eq:dmpk-x}
 \frac{\partial P}{\partial s} =
   \frac{1}{4(N-1)}\sum_{a=1}^{m} \frac{\partial}{\partial x_{a}}
   \left(  \frac{\partial P}{\partial x_{a}}
         + P \frac{\partial \Omega}{\partial x_{a}}
   \right)
\end{equation}
with
\begin{equation}
 \Omega = - \ln \biggl( J \prod_{a=1}^{m} \sinh 2 x_{a} \biggr) .
\end{equation}
In the limit of $s/N = L/Nl \to \infty$,
we expect that the variables $x_{a}$ ($1 \le a \le m$) are much larger
than unity and are widely separated.
We assume that $1 \ll x_{1} \ll x_{2} \ll \cdots \ll x_{m}$.
Thus, we can approximate that
$\Omega \approx -2\sum_{a=1}^{m}(4a-1)x_{a} + {\rm constant}$.
Substituting this into eq.~(\ref{eq:dmpk-x}), we obtain
\begin{equation}
 \frac{\partial P}{\partial s} =
   \frac{1}{4(N-1)}\sum_{a=1}^{m} \frac{\partial}{\partial x_{a}}
   \left( \frac{\partial P}{\partial x_{a}} - 2(4a-1)P \right) .
\end{equation}
The solution of the above equation is
\begin{equation}
  P(x_{1}, \cdots, x_{m}, s)
  = \left( \frac{\gamma}{2\pi s} \right)^{\frac{m}{2}}
    \prod_{a=1}^{m}
    {\rm e}^{- \frac{\gamma}{2s}
               \left( x_{a} - \frac{l}{\xi_{a}}s \right)^{2}} ,
\end{equation}
where $\gamma = 2(N-1)$ and
\begin{equation}
  \xi_{a}=\frac{\gamma l}{4a-1} .
\end{equation}

We observe that each eigenvalue obeys a Gaussian.
From the resulting distribution, we obtain
$\langle x_{a+1} - x_{a} \rangle = 4s/\gamma$
and $\sqrt{{\rm Var} [x_{a}]} = \sqrt{s/\gamma}$,
where ${\rm Var} [A] = \langle A^{2} \rangle - \langle A \rangle^{2}$.
This denotes that the variables are widely separated
as long as $s/\gamma = L/\gamma l \gg 1$.
Note that $\langle x_{1} \rangle = 3s/\gamma$.
Thus, when $3s/\gamma = 3L/\gamma l \gg 1$,
the deviation of the conductance from the quantized value $G_{0} = 4e^{2}/h$
is dominated by $x_{1}$ as
$\delta G \equiv G - G_{0} \sim G_{0} \times 8 {\rm e}^{-2x_{1}}$.
Taking the ensemble average, we obtain
\begin{align}
  - \big\langle \ln ( \delta G/G_{0} ) \big\rangle
      = \frac{3}{N-1} \left( \frac{L}{l} \right) ,
             \\
         {\rm Var} \left[ \ln ( \delta G/G_{0} ) \right]
      = \frac{2}{N-1} \left( \frac{L}{l} \right) .
\end{align}
We estimate the decay length $\xi$ of the typical conductance by identifying
$\exp \langle \ln (\delta G/G_{0}) \rangle = \exp (-2L/\xi)$.
We find that
\begin{equation}
  \xi = \frac{2}{3} (N-1) l = \frac{4}{3} m l.
\end{equation}
The decay length corresponds to the localization length
in ordinary quantum wire systems.
However, wave functions in our system are not localized
due to the presence of the one perfectly conducting channel,
so we do not call $\xi$ the localization length.
The average of $\delta G/G_{0}$ is calculated as
\begin{align}
     \label{eq:average-G}
 \langle \delta G/G_{0} \rangle
 & = 8 \sqrt{\frac{\gamma}{2\pi s}} \int_{0}^{\infty} {\rm d}x {\rm e}^{-2x}
     {\rm e}^{- \frac{\gamma}{2s}(x-3s/\gamma)^{2}}
     \nonumber \\
 & = 8 {\rm e}^{- \frac{2}{N-1}\frac{L}{l}} .
\end{align}
This result denotes that $\langle \delta G/G_{0} \rangle$ decays more slowly
than $\exp \langle \ln (\delta G/G_{0}) \rangle$.
Equation~(\ref{eq:average-G}) is also obtained from eq.~(\ref{eq:tp}).
If we set $p = 1$, eq.~(\ref{eq:tp}) is reduced to
\begin{equation}
 (N-1) \frac{\partial \langle \Gamma \rangle}{\partial s}
  = - \langle \Gamma^{2} \rangle + \langle \Gamma_{2} \rangle .
\end{equation}
Since the deviation of the conductance is dominated by $x_{1}$
when $L/\gamma l \gg 1$, we can approximate that
$\Gamma \sim 1+2T_{1}$ and $\Gamma_{2} \sim 1+2T_{1}^{2}$.
Substituting them into the above equation
and retaining the lowest order terms with respect to $T_{1}$, we obtain
$2(N-1)\partial \langle T_{1} \rangle/\partial s = - 4 \langle T_{1} \rangle$.
This is easily solved as
$\langle T_{1} \rangle \propto {\rm e}^{- \frac{2}{N-1}\frac{L}{l}}$,
which is equivalent to eq.~(\ref{eq:average-G}).

We compare our result with that for the case
where the number of conducting channels is even (i.e., $N = 2m$).
It should be emphasized that there is no perfectly conducting channel
in the even-channel case.~\cite{ando2}
Although the even-channel case is not realized in CNs,
the comparison provides us physical insight into the role of
the perfectly conducting channel.
If we adapt our argument to the even-channel case,
it is reduced to that for the ordinary symplectic class~\cite{beenakker}
with $m$ channels.
This is consistent with Suzuura and Ando's argument~\cite{suzuura}
that a two-dimensional graphite sheet belongs to the symplectic class
when the potential range of scatterers is larger than the lattice constant.
In the symplectic class with $m$ channels,
the localization length is $\xi_{\beta=4} = (4 m - 2) l$.
We observe that the decay length $\xi$ is a factor $1/3$ smaller than
$\xi_{\beta=4}$ in the large-$m$ limit.
This reduction should be attributed to the presence of
the perfectly conducting channel, for which $x_{N} = 0$.
The repulsion between $x_{N} = 0$ and the dominant eigenvalue $x_{1}$ tends
to increase $x_{1}$, resulting in the reduction of $\xi$.

Thus far, we have treated CNs in the absence of a magnetic field.
If a magnetic field is applied, the symmetry relations given in
eqs.~(\ref{eq:symmetry1}) and (\ref{eq:symmetry2}) no longer hold.
Consequently, the system falls into the unitary class.
This transition is characterized by
the disappearance of the perfectly conducting channel
and the increase in decay length ($\xi \to \xi_{\beta=2} \equiv 2Nl$).

In summary, we have studied the quantum electron transport in metallic
carbon nanotubes with several conducting channels when the potential
range of scatterers is larger than the lattice constant.
Considering the unique scattering symmetry observed in carbon nanotubes,
we derived the DMPK equation for
the distribution of the transmission eigenvalues.
With increasing system length $L$,
the conductance decreases towards the quantized value $G_{0}=4e^{2}/h$.
We calculated several statistical averages which characterize
the asymptotic behavior of the conductance,
based on the DMPK equation.
It is shown that
the length scale for the exponential decay of the typical conductance
is reduced due to the presence of the perfectly conducting channel.
The role of a magnetic field is discussed briefly.

\end{document}